\documentclass[twocolumn,showpacs,prl]{revtex4}% Physical Review B
\usepackage{bm}
\usepackage{epsfig}
\newcommand{\nix}[1]{}

\begin{document}

\title{
Circular Photogalvanic Effect at Inter-Band Excitation
\\in Semiconductor Quantum Wells }
\author{V.V.~Bel'kov$^{1,2}$, S.D.~Ganichev$^{1,2}$, Petra~Schneider$^1$, C.~Back$^1$, M.~Oestreich$^3$,
J.~Rudolph$^3$, D.~H{\"a}gele$^3$, L.E.~Golub$^2$,
W.~Wegscheider$^{1,4}$, W.~Prettl$^1$}
\affiliation{$^1$Fakult\"{a}t f{\"u}r Physik, Universit{\"a}t
Regensburg, 93040 Regensburg, Germany}
\affiliation{$^2$A.F.~Ioffe Physico-Technical Institute of the
RAS, 194021 St.~Petersburg, Russia}
\affiliation{$^3$ Institut f{\"u}r Festk{\"o}rperphysik,
Universit{\"a}t  Hannover, 30167 Hannover,
 Germany}
\affiliation{$^4$ Walter Schottky Institute, TU Munich, 85748
Garching, Germany}

%\date{\today}

\begin{abstract}
We observed a circular photogalvanic effect (CPGE) in GaAs quantum
wells at inter-band excitation. The spectral dependence of the
CPGE is measured together with that of the polarization degree of
the time resolved photoluminescence. A theoretical model
takes into account spin splitting of conduction and valence bands.
\end{abstract}
%\pacs{73.21.Fg, 72.25.Fe, 78.67.De, 73.63.Hs}

\maketitle

Spin photocurrents generated by excitation with circularly
polarized  radi\-ation in quantum wells have attracted
considerable attention in the recent decade~\cite{review2003spin}.
Se\-veral mechanisms of electric currents driven by optically
generated spin polarization are observed in zinc-blende-structure
based bulk semiconductors and quantum wells (QWs). Among these
effect are inhomogeneous spin orientation induced currents in bulk
GaAs~\cite{Averkiev83p393,Bakun84p1293}, the circular
photogalvanic effect (CPGE) and the spin-galvanic effect in
QWs~\cite{PRL01,Nature02}, the photovoltaic effect in $p-n$
junctions~\cite{Zutic01p1558,Zutic02p066603} and currents due to
quantum interference of one- and two-photon
excitations~\cite{Bhat00p5432,Stevens02p4382,vanDriel03}. Except
CPGE in QWs, all other spin photocurrents  have been observed at
optical excitation across the band gap of the semiconductor. CPGE~\cite{Ivchenko78p74,Belinicher78p213,Asnin78p74,sturman,book}
is caused in zinc-blende structure based QWs  by homogeneous
optical spin orientation of carriers~\cite{PRL01}.  This effect
should also occur at inter-band excitation~\cite{PRL01,Golub03p2},
but so far has been detected only at intra-band transitions by
excitation with infrared radiation. In the present work we report
on the first observation of the CPGE at inter-band excitation in
GaAs QWs.

\begin{figure}
\begin{center}
\includegraphics[width=7cm]{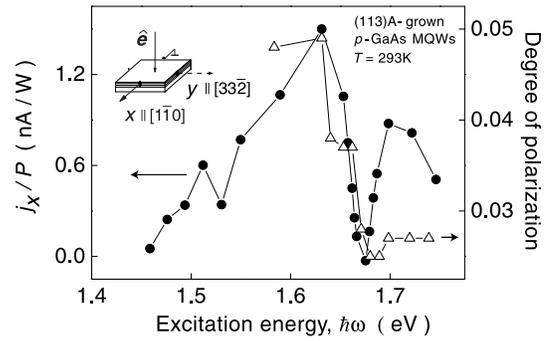}
\end{center}
\caption{Photocurrent in QWs normalized by $P$ and a spectrally
integrated polarization degree of photoluminescence as a function
of the excitation photon energy $\hbar \omega$.  The  inset  shows
the geometry of the experiment. Normal incidence of radiation on
$p$-type (113)A- grown GaAs/AlGaAs QWs (symmetry class C$_s$).}
\label{fig1}
\end{figure}

The experiments were  carried out on (113)A-oriented
molecular-beam-epitaxy (MBE) grown  $p$-type
GaAs/Al$_{0.32}$Ga$_{0.68}$As structure
 with 20 QW of  15~nm widths.  The free hole density
 in the sample was $2\cdot 10^{11}$~cm$^{-2}$ and the mobility
 was about $5\cdot
10^5$~cm$^2$/Vs at 4.2~K. The sample edges were oriented along the
[1$\bar{1}$0]- and [33$\bar{2}$]- directions. Two pairs of ohmic
contacts   were centered along opposite sample edges pointing in
the directions $x\parallel $ [1$\bar{1}$0] and $y
\parallel $ [33$\bar{2}$] (see Fig.~\ref{fig1}).
The sample  belongs to the symmetry class $C_s$ which allows the CPGE at
normal incidence of the radiation~\cite{review2003spin}. For
optical inter-band excitation a cw-Ti:sapphire laser and  pulsed
Ti:sapphire laser were used providing radiation of wavelength in
the range between 0.7~$\mu$m  and 0.85~$\mu$m. The power of the
cw-laser $P$ was about 80~mW. The pulsed laser provided 1~ps pulses
with a repetition rate of 80~MHz and an average power of about
$100$~mW. One of the main features of the CPGE is that the
photocurrent caused by spin polarization is proportional to the
helicity of the incident light $P_{circ}=
(I_{\sigma_+}-I_{\sigma_-})/(I_{\sigma_+} + I_{\sigma_-})$, where
$I_{\sigma_+}$ and $I_{\sigma_-}$ are intensities of right-
($\sigma_+$) and left-handed ($\sigma_-$) polarized radiation.
Therefore, the sign of the current changes upon switching  from
right to left circular polarization. This allows to distinguish the
helicity dependent photocurrent from helicity independent currents
like that of the Dember effect, photovoltaic effects at contacts
and Schottky barriers. The linearly polarized laser beam was
transmitted through a photoelastic modulator which yields  a
periodically oscillating polarization between $\sigma_+$ and
$\sigma_-$.  The photocurrent $j_x$ was measured in the unbiased
structures at room temperature via the voltage drop across a
50~$\Omega$ load resistor in a closed circuit configuration. The
signal was recorded by a lock-in amplifier in phase with the
photoelastic modulator. In addition we carried out polarization
and time resolved photoluminescence measurements using a synchroscan
streak camera with a temporal and spectral resolution of 7~ps and
1~nm, respectively~\cite{Oestreich2002p285}.

\begin{figure}
\begin{center}
\includegraphics[width=7cm]{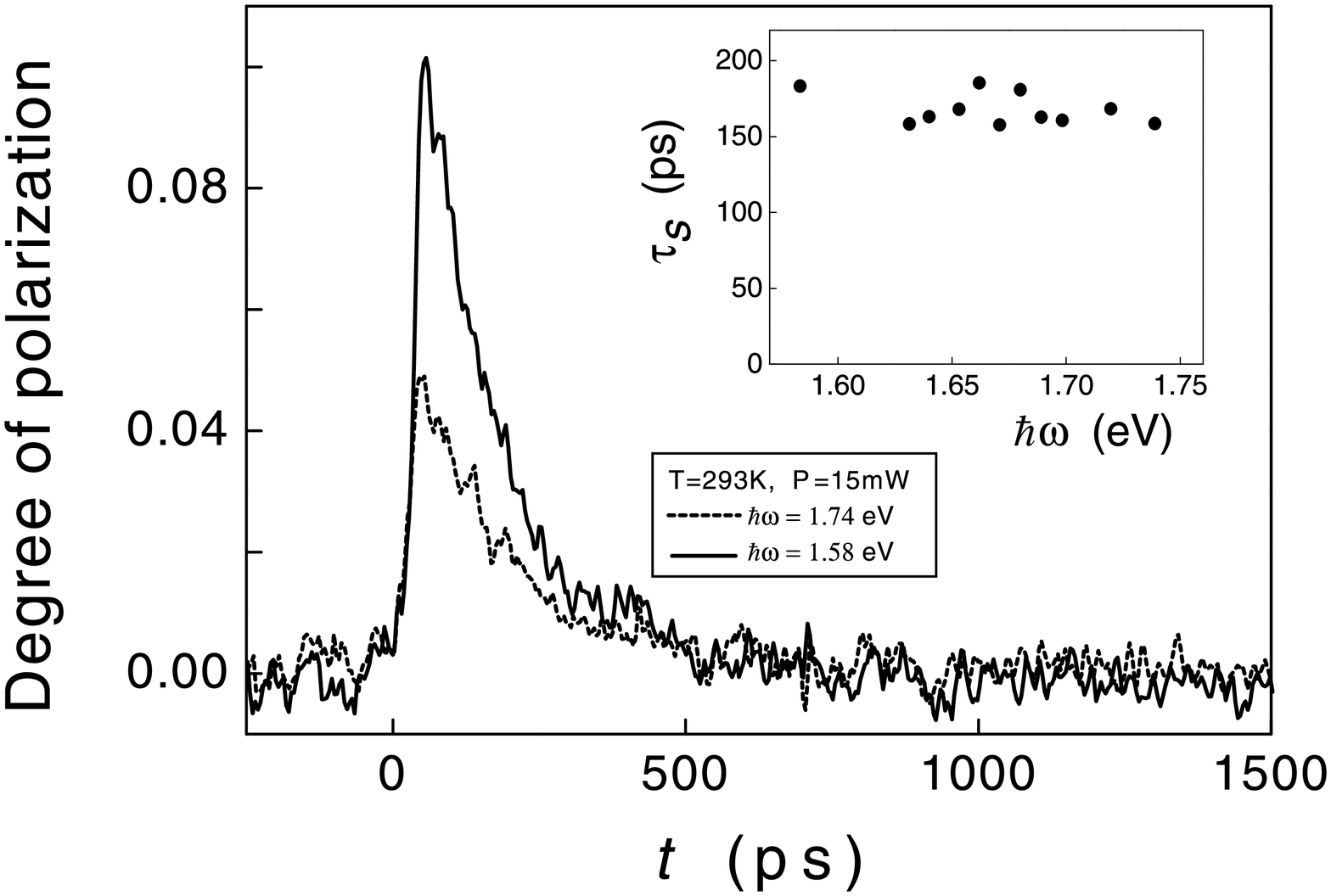}
\end{center}
\caption{ Kinetics of the photoluminescence signal for two various
photon energies. The inset shows the spectral behaviour of the
relaxation rate of electron spins.} \label{fig2}
\end{figure}

Illuminating our QW structure with polarization modulated
radiation at normal incidence we observed a current signal in
$x$-direction  for both cw and pulsed excitations. In the
perpendicular $y$-direction  no helicity dependent current has
been detected. Fig.~\ref{fig1} shows the spectral dependence of
the photocurrent and the spectrally integrated polarization degree
of photoluminescence. Both, photocurrent and polarization degree
have peculiarities in their spectral dependences. In particular,
there is a deep drop of both values at the photon energy $\hbar
\omega = 1.67$~eV. While the photocurrent exhibits a distinct
minimum, the polarization degree sharply drops (by a factor of
two) with increasing energy. In the next step we studied the
temporal dependence of the polarization degree on the excitation
photon energy. Fig.~\ref{fig2} shows that the relaxation time of
electron spins $\tau_s$ is about 170~ps and is independent of
$\hbar \omega$.

On the macroscopic  level the CPGE can be described by the
following  phenomenological expression~\cite{book}:
\begin{equation}
j_{\lambda} = \sum_{\mu}\gamma_{\lambda \mu}\:i (\mbox{\boldmath$
E$} \times {\mbox{\boldmath $E$}}^* )_{\mu} =
\sum_{\mu}\gamma_{\lambda \mu} \hat{e}_\mu \:E_0^2 P_{circ}\:,
\label{equ10}
\end{equation}
where {\boldmath$j$} is the  photocurrent density,
{\boldmath$\gamma$} is a second rank pseudo-tensor, {\boldmath$E$}
is the complex amplitude of the electric field of the
electromagnetic wave, and $E_0$,   {\boldmath$\hat{e}$}  are the
electric field amplitude and the unit vector pointing in the
direction of light propagation, respectively. For QWs of $C_s$
symmetry, as investigated here, a photocurrent  at normal
incidence, with radiation propagating along $z \parallel [113]$,
is given by
\begin{equation}
j_x = (  \gamma_{x z } \hat{e}_{z}  ) E^2_0
P_{circ}\:,\,\,\,\,\,\,\,\,\,\,j_y = 0 \:. \label{equ15}
\end{equation}
The current  flows in $x$-direction which is perpendicular to the
mirror reflection plane of $C_s$ symmetry. Eq.~(\ref{equ15})
describes the experimental observation, in particular, the absence
of a helicity dependent current in $y$-direction.

\begin{figure}
\begin{center}
\includegraphics[width=4cm]{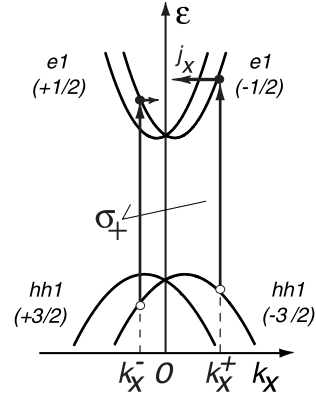}
\end{center}
\caption{Microscopic picture of  spin orientation induced CPGE at
direct transitions in C$_s$ point group taking into account the
splitting of subbands in {\boldmath$k$}-space. $\sigma_+$
excitation induces direct transitions (vertical arrows) (a)
between valence and conduction band (from $hh${\it 1} ($s = -3/2$)
to $e${\it 1} ($s=-1/2$)). Spin splitting together with optical
selection rules results in an unbalanced occupation of the
positive $k_x^+$ and negative $k_x^-$ states yielding a spin
polarized photocurrent. For $\sigma_-$ excitation both the spin
orientation of the charge carriers and the current direction get
reversed. Horizontal arrows indicate the current due to an
unbalance of carriers. Currents are  shown for  conduction band
only.}
\label{fig3}
\end{figure}

Fig.~\ref{fig3} shows the microscopic picture of the spin
orientation induced circular photogalvanic effect at interband
excitation close to the band edge. We assume direct inter-band
transitions in a QW of $C_s$ symmetry. For the sake of simplicity
we discuss a band structure consisting only of the lowest
conduction subband $e${\it 1} and the highest heavy-hole subband
$hh${\it 1}.   The energy dispersion in the conduction band is
given by
$$\varepsilon_{e{\it 1}, \pm 1/2}(\mbox{\boldmath$k$}) = \hbar^2
k^2/2 m_{e} \pm \beta_{e{\it 1}} k_{x} + \varepsilon_g$$
and in the valence band by
$$\varepsilon_{hh{\it 1}, \pm 3/2}(\mbox{\boldmath$k$}) =
-[\hbar^2 k^2/2 m_{hh{\it 1}} \pm \beta_{hh{\it 1}} k_{x }],
$$
where $\varepsilon_g$ is the QW energy gap,  $\beta_{e1}$ and
$\beta_{hh1}$ are components of second rank pseudo-tensor
responsible for the removal of spin degeneracy in the first electron
and the first heavy-hole subbands, respectively.

For absorption of circularly polarized radiation of photon energy
$\hbar\omega$, momentum  and  energy conservation allow
transitions only for two  values of $k_x$. Due to the selection rules
the optical transitions occur from $s=-3/2$ to $s=-1/2$ for right
handed circular polarization ($\sigma_{+}$) and from $s = 3/2$ to
$s = 1/2$ for left handed circular polarization ($\sigma_{-}$).
Here $s$ are the spin quantum numbers of the electron states. The
corresponding transitions for e.g. $\sigma_{+}$ photons  occur at
\begin{equation}
k_{x }^{\pm} =  \frac{\mu}{\hbar^2} (\beta_{e{\it 1}}+\beta_{hh{\it
1}} )\pm \sqrt{ \frac{\mu^2}{\hbar^4} (\beta_{e{\it 1}}
+\beta_{hh{\it 1}} )^2 + \frac{2 \mu}{\hbar^2}\left( \hbar \omega
- \varepsilon_g \right)}, \label{equ9}
\end{equation}
and are shown in Fig.~\ref{fig3} by the solid  vertical arrows.
Here $\mu = (m_{e}\cdot m_{hh{\it 1}})/(m_{e}+m_{hh{\it 1}})$ is
the reduced mass. The `center of mass' of these transitions is
shifted from the point \mbox{$k_{x}=0$} by $(\beta_{e{\it
1}}+\beta_{hh{\it 1}}) (\mu / \hbar^2)$. Thus the sum of the
electron velocities in the excited states in the conduction band,
\nix{
\begin{eqnarray}
  v_{e{\it 1}} &=& {\hbar \over  m_{e}} \: (k_{x }^{-} + k_{x }^{+}-2k_x^{min})  \nonumber \\
  &=& {2 \over \hbar} \: {\beta_{hh{\it 1}}m_{hh{\it
1}}-\beta_{e{\it 1}}m_{e} \over m_{e} + m_{hh{\it 1}} }, \nonumber
\end{eqnarray}
}
\begin{equation}
      v_{e{\it 1}} = {\hbar \over  m_{e}} \: (k_{x }^{-} + k_{x }^{+}-2k_x^{min})
  = {2 \over \hbar} \: {\beta_{hh{\it 1}}m_{hh{\it
1}}-\beta_{e{\it 1}}m_{e} \over m_{e} + m_{hh{\it 1}} },
\end{equation}
is non-zero. The contributions of $k_{x}^{\pm}$ photoelectrons to
the current do not cancel each other except in the case of
$\beta_{e{\it 1}}m_{e}=\beta_{hh{\it 1}}m_{hh{\it 1}}$ which
corresponds to an equal splitting of the conduction and the
valence band. We note that the group velocity is obtained taken
into account that $k_{x}^{\pm}$ are to be counted from the
conduction subband minima $k_x^{min}$ because the current is
caused by the difference  of the group velocities within the
subband. The same consideration applies for holes in the initial
states in $hh${\it 1}. Consequently, a spin polarized net current
results in the $x$ direction. Changing the circular polarization
of the radiation from $\sigma_{+}$ to $\sigma_{-}$ reverses  the
current because the `center of mass' of these transitions is now
shifted to $-(\beta_{e{\it 1}}+\beta_{hh{\it 1}}) (\mu /
\hbar^2)$.

The microscopic theory of spin orientation induced CPGE in QWs at
inter-band excitation was developed in~\cite{Golub03p2}. We
consider the asymmetry of the momentum distribution of electrons
excited under direct inter-band optical transitions in $p$-doped
$(113)$-grown QWs of $C_s$ symmetry. The photocurrent density is
given by
\begin{equation}
\label{j} {\bm j} = e \sum_{\bm k} {\rm Tr} \left[ {\bm v}^{(e)}
({\bm k}) \: \tau^{(e)} \: \dot{\rho}^{(e)}({\bm k}) + {\bm
v}^{(h)} ({\bm k}) \: \tau^{(h)} \: \dot{\rho}^{(h)}({\bm k})
\right] \:,
\end{equation}
where $e$ is the elementary charge, ${\bm v}^{(e, h)} ({\bm k})=
(1/\hbar)
\partial \varepsilon^{(e,h)}_{\nu s} ({\bm k}) / \partial {\bm k}$
are the group velocities  of electrons and holes, $\tau^{(e,h)}$
are the momentum relaxation times in the subbands, and $\dot{\rho}^{(e,h)}$ are
their non-equilibrium generation rates. The expressions for them
have the form
\begin{equation}
\label{rho} \dot{\rho}^{(e,h)}_{nn'} = {\pi \over \hbar} \sum_l
M_{nl}M_{ln'} [\delta (E_n + E_l - \hbar \omega) + \delta (E_{n'}
+ E_l - \hbar \omega)] \:.
\end{equation}
Here $M_{nl}({\bm k})$ is the inter-band optical matrix element,
%proportional to
%the amplitude of the electromagnetic field
%$E_0$,
and the indices $n, n', l$ enumerate the electron and hole
subband spin states involved into the transitions. The CPGE
current excitation spectrum is given by
\begin{equation}\label{jx}
    j_x(\omega)  = \sum_{\nu, \nu'} \: [ \beta_{e\nu} F_{\nu \nu'} (\omega) + \beta_{h \nu'}
    \tilde{F}_{\nu \nu'}(\omega)],
\end{equation}
where $\beta_{e\nu}$, $\beta_{h\nu'}$ are the spin-splitting
parameters for $\nu^{th}$ electron and $\nu'^{th}$ hole subbands of
size-quantization. The spectral functions $F_{\nu \nu'}(\omega)$
and $\tilde{F}_{\nu \nu'}$ are calculated in~\cite{Golub03p2}. The
most important result of the microscopic theory is that both the
initial and final states of the carriers involved in the optical
transition contribute to  the circular photogalvanic current with
different strength and directions. The partial currents are
proportional to the  group velocity being dependent
on~{\boldmath$k$}, the momentum relaxation times~$\tau_e$ and
~$\tau_h$, and the occupation of the initial states. Therefore the
sign of the total current depends on the details of experimental
conditions and may even change by varying the radiation frequency,
temperature etc.

The photocurrent is due to carrier spin polarization, thus the
coincidence of spectral peculiarities in both the CPGE and the
circular polarization degree is natural. The reason for the
peculiarities in both spectra is a sharp  dependence of spin
polarization on the wavevector for the direct inter-band
transitions. Close to the band edge, the transitions obey the
simple selection rules and occur from pure $\pm 3/2$ heavy-hole
states as assumed in the model above. Increase of the excitation
energy shifts the transitions to higher wavevectors at which the
selection rules are lifted. This is caused by the heavy-light hole
mixing in QWs leading to a strong dependence of carrier spin
orientation on the excitation energy theoretically considered
in~\cite{Merkulov1991}. On the other hand, transitions from
 hole subbands to electron continuum are possible at the excitation energy
 $\hbar\omega \gtrsim $ 1.7~eV. Such process is known as 'photoionization' of QWs~\cite{Feng}.

In conclusion we have observed the CPGE in QW structures at
inter-band excitation for the first time. The strength of the CPGE
signal correlates with the degree of spin polarization and is well
described by phenomenological and microscopical theories.

 Acknowledgements: We thank E.L.Ivchenko for helpful discussions.
Authors gratefully acknowledge financial support by the Deutsche
Forschungsgemeinschaft (DFG) and the INTAS.

\end{document}